\documentclass[prb,aps,amsmath,amssymb,reprint]{revtex4-1}
\usepackage{graphicx}
\usepackage{dcolumn}
\usepackage{bm}
\newcommand{\gs}{$\gamma^{\prime}$}

\newcommand{\GMM}{$\alpha^{\prime\prime}\mathrm{-Fe_{16}N_{2}}$}

\begin{document}

\title{Effect of dopants on thermal stability and self-diffusion in iron nitride thin films}

\author{Akhil Tayal, Mukul Gupta}
\email{mgupta@csr.res.in/dr.mukul.gupta@gmail.com}
\author{Ajay Gupta}

\affiliation{UGC-DAE Consortium for Scientific Research,
University Campus, Khandwa Road, Indore 452 001, India}

\author {M. Horisberger$^1$ and Jochen Stahn$^2$} \affiliation{$^1$LDM,$^2$LNS, Paul Scherrer Institut, CH-5232 Villigen
PSI, Switzerland}

\author {Kai Schlage and H.-C. Wille}
\affiliation{Deutsches Elektronen-Synchrotron DESY, Notkestrasse
85, D-22607 Hamburg, Germany}


\begin{abstract}

We studied the effect of dopants (Al, Ti, Zr) on the thermal
stability of iron nitride thin films prepared using a dc magnetron
sputtering technique. Structure and magnetic characterization of
deposited samples reveal that the thermal stability together with
soft magnetic properties of iron nitride thin films get
significantly improved with doping. To understand the observed
results, detailed Fe and N self-diffusion measurements were
performed. It was observed that N self-diffusion gets suppressed
with Al doping whereas Ti or Zr doping results in somewhat faster
N diffusion. On the other hand Fe self-diffusion seems to get
suppressed with any dopant of which heat of nitride formation is
significantly smaller than that of iron nitride. Importantly, it
was observed that N self-diffusion plays only a trivial role, as
compared to Fe self-diffusion, in affecting the thermal stability
of iron nitride thin films. Based on the obtained results effect
of dopants on self-diffusion process is discussed.
\end{abstract}

\date{\today}
\maketitle

\section{Introduction}
\label{Sec_Intro}

Iron nitride (Fe-N) compounds exist in a variety of phases having
distinct crystal structure and magnetic
properties.~\cite{Jack1951} These compounds have gained a special
attention for their numerous technological applications such as in
tribological coatings, magnetic memory devices, high frequency
read write heads
etc.~\cite{Navio.PRB08,Gallego.PRB04,PRB:Fe4N:SpinPol,SM:Tribological,ding:JAP97,IEEE:Kryder:Data,JAP:Georgieva:SoftMag,Jiang:JPCM94:Fe16N2,Sawada:PRB94}
Different Fe-N phases can be formed by varying the nitrogen
concentration in Fe.~\cite{PD:FeN:2010} When N concentration is
less than 25at.\%, Fe-N compounds are ferromagnetic and their
Curie temperature is above room temperature. Strikingly, within
the composition range 0$<$N$\leq$11at.\%, Fe-N compounds posses
interesting magnetic properties and major phases formed are
$\alpha$-Fe, Fe$_8$N and
\GMM.~\cite{Schaaf.PMS.2002,Borsa.HI.2003} In this composition
range, N atoms get mainly occupied at random-interstitial sites
within the $bcc$ Fe lattice. This creates some distortion in the
lattice and results in nanocrystallization, which improves
soft-magnetic properties and favors magnetic
anisotropy.~\cite{Gupta:PRB05,Das:PRB:2007,Ranjeeta_PRB12,Ranu.APL.2008}
Around the saturation concentration i.e. $\sim$11at.\% instead of
random occupancy, N atoms get perfectly ordered to form \GMM~phase
having $bct$ structure.~\cite{Kim:APL:1972} At around 20at.\% of N
\gs-Fe$_4$N phase is formed having $fcc$ structure. These Fe-N
phases have tremendous applications due to their ingenious
magnetic
properties.~\cite{Navio.PRB08,Gallego.PRB04,PRB:Fe4N:SpinPol,Navio:APL:2009,Jiang:JPCM94:Fe16N2,Sawada:PRB94,Sun:JPCM95:Fe16N2}
However, weak Fe and N bonding results in poor thermal stability
of Fe-N
compounds.~\cite{Gupta_PRB02,Gupta:AM:2009,Ding:IEEE:2006,gupta:JAP2011,Chechenin:JPCM:2003}
It causes a severe limitation for Fe-N compounds to succeed as a
potential candidate for device applications.

Recently, it was shown that doping of a third element, say X (e.g.
Al, Ti, Zr, Ta, etc.) results in superior thermal stability of
Fe-N thin
films.~\cite{Chechenin:JPCM:2003,Hasegawa:JMSJ:1990,Wang:JPCM99:FeCoN,Ishiwata:JAP:1991,Ono:JAP:1993,Takeshima:JAP:1993,
Qiu:JAP:1994,Roozeboom:JAP:1995,Viala:JAP:1996,Wang:JPCM:1997,Varga:JAP:1998,
Chen:JAP:2000,Liu:APL:2000,Chezan:PSSA:2002,Rantschler:JAP:2003,Liu:JAP:2003,Kazmin:TPL:2005,
Das:PRB:2007,Sangita:PRB:2008,FengXu:JAP:2011,RG:JAP12} These
dopants either gets dissolved substitutionally into the Fe lattice
or form a solid solution with Fe, which also creates some lattice
distortion. Importantly, the qualifying thermodynamical variables
that are required to enhance the thermal stability are high
affinity ($f$) and low heat of formation ($\Delta$H) of X-N as
compared to Fe-N. It has been established in the literature that
high X-N bonding may suppress N diffusion to enhance the thermal
stability of resultant Fe-X-N thin
films.~\cite{Kopcewicz:JAP:1995} Although, no direct diffusion
measurements are available to support such arguments, besides, the
role of Fe self-diffusion is also not yet clearly established.
Apart from it, on the basis of this assumption, it is expected
that X with the highest affinity for N and the smallest $\Delta$H
should be an ideal choice. However, numerous studies that were
made using various dopants gives a diverse picture on their role
in affecting the structure, magnetic properties and thermal
stability that makes the choice of dopants rather
arbitrary.~\cite{Chechenin:JPCM:2003,Wang:JPCM99:FeCoN,
Das:PRB:2007,Liu:APL:2000,Viala:JAP:1996,Varga:JAP:1998} In some
studies, it was observed that doping of Al or Ta improves the soft
magnetic properties. Even so, thermal stability is found to get
better with Al addition as compared to
Ta.~\cite{Liu:APL:2000,FeTaN:Soft:JAP} It was claimed that large
atomic size of Ta results in a lower diffusion barrier for the
interstitial nitrogen. On the contrary, Chechenin $et~al.$ found
that nitrogen desorption get reduced by addition of Zr, although
the atomic size of Zr and Ta is almost
equal.~\cite{Chechenin:JPCM:2003} Similarly, addition of Ti have
found to affect various magnetic properties and significantly
enhance the thermal stability of Fe-N thin films, even through its
atomic size is larger than Fe.~\cite{JPCM:97:Wang,Das:PRB:2007}

Above scenario suggests that beside the thermodynamic parameters
$\Delta$H and $f$, atomic size of X may play a decisive role in
improving the thermal stability of Fe-N thin films.
Table~\ref{tab:table1} compares these parameters for the dopants
mostly used in Fe-X-N thin
films.~\cite{Tessier_SSS00,Kopcewicz:JAP:1995,Evans_PhD,Clementi:JCP:1967}
From table~\ref{tab:table1}, it can be seen that Zr and Al are on
extreme ends with reference to their affinity and atomic size.
Although their heat of nitride formation is somewhat similar,
still significantly lower as compared to magnetic Fe-N. Therefore,
it will be interesting to compare the effect of Al and Zr doping
on thermal stability and atomic diffusion. To the best of our
knowledge this is a first study correlating the effect of nitrogen
and iron self-diffusion and thermal stability in Fe-N thin films.
To obtain a conclusive picture, N self-diffusion measurements were
also performed on Ti doped sample, as Ti doping has also been
extensively used in
literature.~\cite{JPCM:97:Wang,RG:JAP12,TSF:Tayal:13}

\begin{table}[!hb] \center
\vspace{-5mm} \caption{\label{tab:table1} Atomic radius ($r$),
heat of formation ($\Delta$H) and affinity ($f$) of X-N with
respect to Fe-N.}
\begin{tabular}{lccc}
\hline \hline Element&$r$(pm)& $\Delta$H (kJ mol$^{-1}$) & $f$ \\
\hline
Zr&206&-360&-0.63 \\
Ti&176&-338&-0.53 \\
Ta&200&-237&-0.032 \\
Al&118&-321&-0.028 \\ \hline Fe&146&-10&--\\ \hline
\end{tabular}
\vspace{-2mm}
\end{table}

\section{Experimental}
\label{Sec_Exp}

Fe-X-N thin film samples with X = Al, Ti or Zr, were deposited
using a dc-magnetron sputtering technique simultaneously on
Si(100) and float glass substrates at room temperature. Pure Fe,
[Fe+Al], [Fe+Ti] and [Fe+Zr] composite targets were sputtered
using a mixture of N$_2$ (1.5\,sccm) and Ar(8.5\,sccm) gases. To
get the desired composition of Al, Ti and Zr, relative coverage of
Al, Ti or Zr on Fe target was varied taking into account their
relative sputter yields. Composition of deposited samples was
measured using energy dispersive X-ray analysis and secondary
neutral mass spectroscopy and comes out to be 5.6($\pm1.2$)at.\%
for Al , 4.1($\pm1.4$)at.\% for Ti, and 3.2($\pm1.3$)at.\% for Zr.
The composition of nitrogen was measured with secondary ion mass
spectroscopy using a reference sample of known composition and
comes out to be $\sim$11at.\%($\pm1$) in all the samples. This was
also confirmed using conversion electron M\"{o}ssbauer
spectroscopy in the un-doped sample. A base pressure of about
$1\times10^{-7}$\,mbar was achieved prior to the deposition.
During the deposition, partial pressure in the chamber was about
$4\times10^{-3}$\,mbar. More details about the deposition system
are given elsewhere.~\cite{Gupta:PRB05,gupta:JAP2011}

Multilayer samples with nominal structure:
substrate$\mid$[$^{nat}$Fe-X-N(6\,nm)$\mid^{57}$Fe-X-N(6\,nm)]$_{\times10}$
and
substrate$\mid$[Fe-X-$^{nat}$N(9\,nm)$\mid$Fe-X-$^{15}$N(9\,nm)]$_{\times25}$
were prepared for Fe and N self-diffusion measurements using
neutron reflectivity, with X = 0, Al, Ti, and Zr. Here `$nat$' is
indicates isotopes of Fe and N obtained in natural abundance.
Isotope enrichment of $^{57}$Fe layers exceeds to about 95\%, and
that of $^{15}$N is about 98\%. For N diffusion measurements with
secondary ion mass spectroscopy a special trilayer structure:
substrate$\mid$[Fe-X-$^{nat}$N(110\,nm)$\mid$Fe-X-$^{15}$N(2\,nm)$\mid$Fe-X-$^{nat}$N(110\,nm)],
with X = 0, Al, Zr was deposited. Such structure is expected to
give a peak when looking at $^{15}$N depth profile. For Fe
diffusion measurements with nuclear resonance reflectivity,
substrate$\mid$[$^{nat}$Fe-N(2.2\,nm)$\mid^{57}$Fe-N(2.2\,nm)]$_{\times10}$
and
substrate$\mid$[$^{nat}$Fe-Al-N(2\,nm)$\mid^{57}$Fe-Al-N(2\,nm)]$_{\times10}$
samples were deposited. All samples were deposited using identical
deposition condition. Samples with X= 0, Al, Zr and Ti are named
as Fe-N, Fe-Al-N, Fe-Zr-N, Fe-Ti-N, respectively through out this
work.

Structural characterization was done using X-ray diffraction (XRD)
on a Bruker D8 Advance X-ray diffractometer with Cu K$\alpha$
X-ray source in $\theta-\mathrm{2}\theta$ geometry. Magnetic
properties were studied using Quantum Design superconducting
quantum interference device-vibrating sample magnetometer (S-VSM)
and conversion electron M\"{o}ssbauer spectroscopy (CEMS).
Self-diffusion measurements of Fe was performed using polarized
neutron reflectivity (PNR) and nuclear resonance reflectivity
(NRR). For N self-diffusion measurements PNR and secondary ion
mass spectrometry (SIMS) techniques were used. The PNR
measurements were carried out on AMOR reflectometer at SINQ, PSI
Switzerland. A magnetic field of about 400\,kA/m was applied to
saturate the samples magnetically. For diffusion measurements only
spin up reflectivity were used. The NRR measurements were
performed using 14.4\,keV radiation at P01 beamline, PETRA III,
DESY, Germany. SIMS measurements were performed on a Hiden
Analytical SIMS Workstation. A base pressure of 8$\times
\mathrm{10}^{-10}$\,mbar was achieved in the SIMS chamber. A beam
of O$_{2}^{+}$ primary ions (energy 5\,keV and current 400\,nA)
was used to sputter samples. During measurements pressure in the
chamber was about 8$\times \mathrm{10}^{-8}$\,mbar. To investigate
the thermal stability and to measure self-diffusivity $ex-situ$
annealing of samples was performed in a separate vacuum chamber.

\section{Results}

\subsection{X-ray diffraction}
\label{Sec_XRD_ANN}

\begin{figure*}\center
\includegraphics [width=100mm,height=80mm] {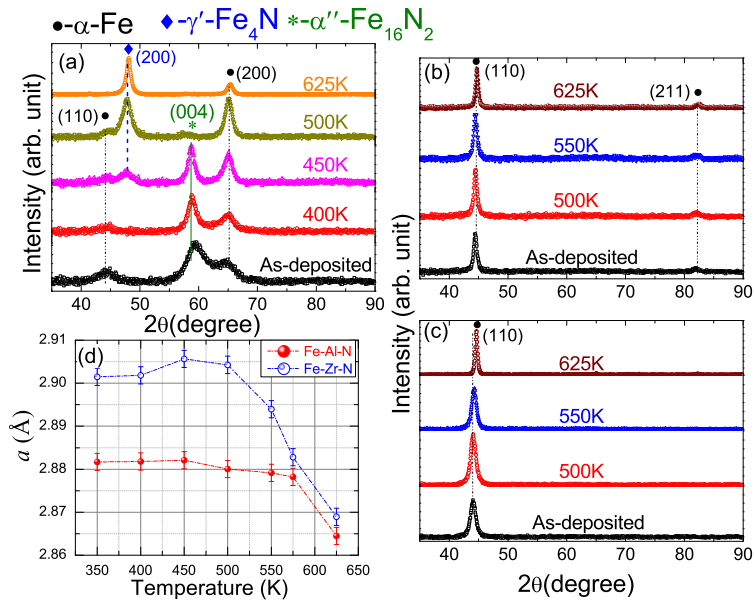}
\vspace {-1mm} \caption{\label{fig:xrd_ann} (color online) XRD
patterns of Fe-N(a), Fe-Al-N(b) and Fe-Zr-N(c) samples in the
as-deposited state and after annealing at various temperatures.
For better comparison scales are vertically translated. Variation
of lattice constant with annealing temperature for Fe-Al-N and
Fe-Zr-N samples(d).}
\end{figure*}

Fig.~\ref{fig:xrd_ann} (a)-(c) shows the XRD patterns of Fe-N,
Fe-Al-N and Fe-Zr-N samples in the as-deposited state and after
annealing at various temperatures. XRD patterns taken at selective
temperatures are shown in the figure. Form
fig.~\ref{fig:xrd_ann}(a) it can be seen that in the as-deposited
state distinct broad peaks corresponding to $bcc~\alpha$-Fe(N) and
\GMM~phase are appearing.~\cite{PRB:Ji:PMA:GMM} It is known that
in these phases nitrogen atoms are occupied within the
interstitial sites of Fe lattice in random and ordered fashion,
respectively.~\cite{Borsa.HI.2003} With annealing up to 450\,K
peak intensity increases and the width reduces indicating an
enhancement in nitrogen ordering. Above this temperature
$\alpha^{\prime\prime}$ phase almost disappears. With further
annealing, intensity of the peak corresponding to \gs-Fe$_4$N
rises implying the growth of \gs~phase. On contrary to this, the
behavior of Fe-Al-N and Fe-Zr-N samples is different. From
fig.~\ref{fig:xrd_ann}(b-c) it can be seen that in the
as-deposited state, only peaks corresponding to $\alpha$-Fe(N)
phase are observed. Annealing of the samples up to 500\,K shows no
change in the XRD pattern. Moreover, no extra peaks corresponding
to any other phases can be seen even after 625\,K. This indicates
that thermal stability of the films gets significantly improved
with doping. From (110) reflection, average crystallite size in
the samples was calculated using Scherrer
formula~\cite{Cullity_XRD} and found to be 15\,nm and 9\,nm for
Fe-Al-N and Fe-Zr-N samples, respectively. Moreover, lattice
constant ($a$) of doped samples in the as-deposited state is
compared with un-doped samples. It was observed that with Al
doping average volume of Fe unit cell get reduced to about 0.5\%,
whereas, in case of Zr it get increased by about 2\%. This results
can be attributed to the varied atomic size of Al and Zr with
respect to Fe. For doped samples it can be seen that the peak
position corresponding to the (110) reflection shifts to higher
2$\theta$ above 500\,K. Fig.~\ref{fig:xrd_ann}(d) shows variation
of the lattice constant with annealing temperature for Fe-Al-N and
Fe-Zr-N samples. Up to 575\,K `$a$' remains almost constant with
Al doping, while it shows a marginal increase above 400\,K and a
steep decrease above 575\,K. Whereas with Zr doping steep decrease
in `$a$' starts already at 500\,K. It signifies that Al doping
results in relatively superior structural stability.

\subsection{Magnetic measurements}
\label{Sec_MH}

\begin{figure}\center
\includegraphics [width=90mm,height=35mm]  {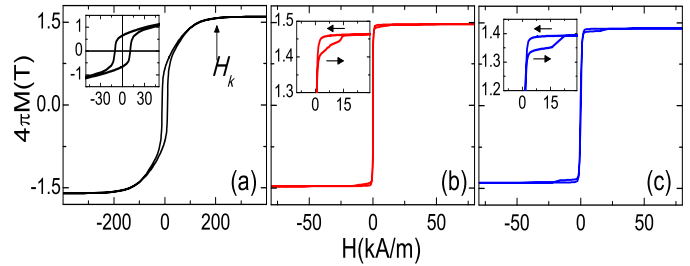}
\caption{\label{fig:mh} (color online) M-H loops of as-deposited
Fe-N(a), Fe-Al-N(b) and Fe-Zr-N(c) thin films. Inset of figure
shows a blown up region near the coercive field. Arrows in (b,c)
show direction of applied magnetic field during M-H measurements.}
\end{figure}

\begin{figure}\center
\includegraphics [width=55mm,height=40mm]  {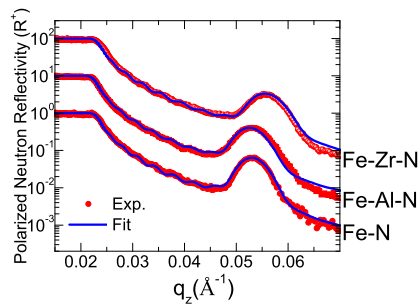}
\caption{\label{fig:pnr15p} (color online) Spin-up reflectivity
patterns of as-deposited
Sub.$\mid$[Fe-N(6.53\,nm)$\mid^{57}$Fe-N(6.53\,nm)]$_{\times 10}$,
Sub.$\mid$[Fe-Al-N(6.55\,nm)$\mid^{57}$Fe-Al-N(6.55\,nm)]$_{\times
10}$, and
Sub.$\mid$[Fe-Zr-N(6.15\,nm)$\mid^{57}$Fe-Zr-N(6.15\,nm)]$_{\times
10}$ thin films. Patterns are vertically translated for clarity.}
\end{figure}

From the XRD results discussed in section~\ref{Sec_XRD_ANN}, it
was found that the as-deposited state of Fe-N thin film comprises
of $\alpha$-Fe(N) and \GMM~phases and in Fe-Al-N and Fe-Zr-N
samples, only $\alpha$-Fe(N) phase is observed. To investigate the
implication of different structure on the magnetic properties, we
did M-H (magnetization-applied magnetic field) measurements.
Fig.~\ref{fig:mh}(a-c) shows M-H loops of the as-deposited
samples. The inset in figures shows a blown up region near the
coercive field. A typical $`transcritical~shape'$ of the M-H loop
(fig.~\ref{fig:mh}(a)) for un-doped sample suggests that the film
exhibits perpendicular magnetic anisotropy (PMA). Recently, Ji
$et~al.$ have reported similar type M-H loops for epitaxial
\GMM~thin films.~\cite{PRB:Ji:PMA:GMM} Here, it was claimed that
observed magnetic anisotropy originates due to magneto crystalline
anisotropy in the system, which originates due to tetragonal
distortion of $bcc$ Fe lattice as N atoms occupy interstitial
positions. Magneto crystalline anisotropy constant ($K_{u}$) can
be calculated using: $K_{u}= H_{k} \times M_{s}/2$
;~\cite{PRB:Ji:PMA:GMM} where $M_{s}$ is saturation magnetization
and $H_{k}$ is anisotropy field which has a value close to $H_{s}$
saturation field as indicated in the fig~\ref{fig:mh}(a). We
obtain $K_{u}$= 5.4$\times10^{5}~\mathrm{Jm}^{-3}$, which is close
to the reported value.~\cite{PRB:Ji:PMA:GMM} Additionally,
obtained values of coercivity ($H_{C}$) = 10\,kA/m and saturation
magnetization (4$\pi M_{S}$) = 1.6\,T. Apart from this, the ratio
of remanence magnetization and $M_{S}$ was found to be 0.4,
indicating PMA. For the Fe-Al-N and Fe-Zr-N samples, M-H loops are
square shaped with a very small value of $H_{C}$($\sim$1\,A/m),
indicating formation of a soft magnetic phase. For Fe-Al-N and
Fe-Zr-N samples the value of 4$\pi M_{S}$ is 1.46\,T and 1.39\,T,
respectively, which are smaller than Fe-N, due to doping of
non-magnetic elements. Further, in the doped samples, an open
region near the saturation field can be seen (as shown in the
inset of the fig.~\ref{fig:mh}(b,c)) indicating presence of some
hard magnetic phase. Existence of hard magnetic phase in the
samples was later confirmed by M\"{o}ssbauer spectroscopy
measurements which are presented in section~\ref{Sec_CEMS}.

Apart from M-H measurements we have also performed PNR
measurements on these samples. It is known that for magnetic thin
films, PNR is a unique technique to measure the magnetic moment
very precisely irrespective of sample dimension. The difference
between the critical edges of spin-up (fig.~\ref{fig:pnr15p}) and
spin-down (not shown) reflectivites provide information about
magnetic moment as samples were magnetically
saturated.~\cite{Blundell_PRB92} Here observed Bragg peak is due
to bilayer periodicity of $\mathrm{^{57}}$Fe/$^{nat}$Fe having
different neutron scattering length contrast. To obtain
information about the bilayer thickness and magnetic moment in the
samples the data are fitted using SimulReflc
software.~\cite{SimulReflec} For Fe-N, Fe-Al-N and Fe-Zr-N samples
bilayer thickness are 13.06\,nm, 13.1\,nm, and 12.3\,nm, and value
of the magnetic moment is 1.8\,$\mu_B$, 1.7\,$\mu_B$, and
1.65\,$\mu_B$, respectively. These values of magnetic moment are
in agreement with those obtained from S-VSM measurements.

\subsection{Conversion electron M\"{o}ssbauer spectroscopy}
\label{Sec_CEMS}

\begin{figure}\center
\vspace{-1mm}
\includegraphics [width=85mm,height=60mm] {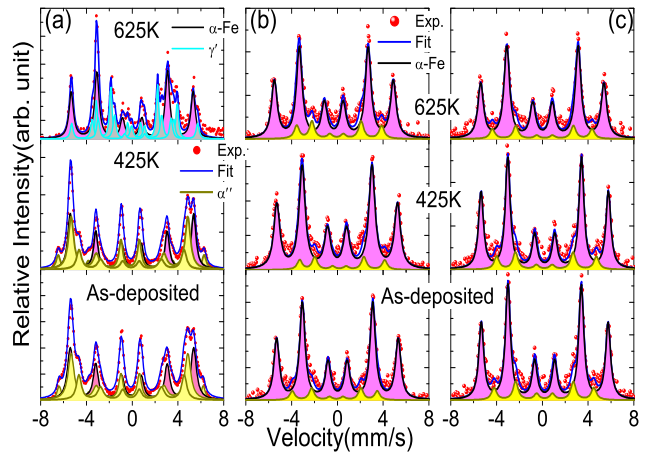}
\vspace {-1mm} \caption{\label{fig:Moss}(color online) CEMS
spectra of Fe-N(a), Fe-Al-N(b), and Fe-Zr-N(c) thin films in the
as-deposited state and after annealing at various temperatures.}
\vspace{3mm}
\end{figure}

\begin{table*}
\caption{\label{tab:table2} Volume fraction of various Fe-N phases
obtained from fitting CEMS spectra. Here `H' stands for a hard
magnetic phase.}
\begin{ruledtabular}
\begin{tabular}{lccc}
Temperature&Fe-N&FeAlN&FeZrN \\ \hline
As-deposited& 59\% (\GMM) + 41\% ($\alpha$-Fe) & 89\% ($\alpha$-Fe) + 11\% (H) & 86\% ($\alpha$-Fe) + 14\% (H) \\
425\,K & 60\% (\GMM) + 40\% ($\alpha$-Fe) & 90\% ($\alpha$-Fe) + 10\% (H) & 86\% ($\alpha$-Fe) + 14\% (H) \\
625\,K & 48\% (\gs) + 52\% ($\alpha$-Fe) & 88\% ($\alpha$-Fe) + 12\% (H) & 88\% ($\alpha$-Fe) + 12\% (H) \\
\end{tabular}
\end{ruledtabular}
\vspace{-5mm}
\end{table*}

M\"{o}ssbauer spectroscopy is a versatile technique to probe the
local structure around a resonant nuclei. From our XRD results,
discussed in the section~\ref{Sec_XRD_ANN}, it was observed that
in the as-deposited state Fe-N films have mixed \GMM~and
$\alpha$-Fe(N) phases. Whereas in Al or Zr doped samples, only
$\alpha$-Fe(N) phase can be seen. Using M\"{o}ssbauer spectroscopy
relative volume fraction of the different Fe-N phases can be
obtained. Fig.~\ref{fig:Moss}(a-c) shows selective CEMS spectra of
Fe-N(a) Fe-Al-N(b) and Fe-Zr-N(c) thin films in the as-deposited
state and after annealing at 425\,K and 625\,K. We fitted the
observed CEMS spectra using NORMOS SITE and DIST
programs.~\cite{Normos:brand:95}

CEMS spectrum of the as-deposited Fe-N sample can only be fitted
assuming four sextet. Three of them corresponding to \GMM~phase
(hyperfine field = 39\,T, 31.6\,T, and 28.5\,T) and remaining to
$\alpha$-Fe (hyperfine field=33\,T), the obtained fitting
parameters correlates-well with previously reported
values.~\cite{Takahashi:Fe16N2} Additionally, area of the fitted
sextets can be used to calculate the relative concentration of
these phases, which are tabulated in the table~\ref{tab:table2}.
As can be seen, after annealing at 425\,K the values are almost
similar to as-deposited sample, but at a higher annealing
temperature of 500\,K (not shown) and 625\,K, \GMM~phase
disappears and \gs phase starts to grow. Observed CEMS results for
the Fe-N sample correlate well with our XRD results. Moreover,
from the relative area under \gs~and $\alpha$-Fe phase above
500\,K, N concentration can be estimated~\cite{Ranjeeta_PRB12}
which comes out to be (11$\pm$2)at.\% which is in agreement with
our SIMS measurements.

In comparison to Fe-N, CEMS spectra of doped samples are
completely different and remains almost similar to the
as-deposited samples even after annealing at 625\,K, as expected
from XRD results. Besides, it was found that CEMS spectra of
Fe-Al-N and Fe-Zr-N samples can only be fitted assuming two
components - one corresponding to $\alpha$-Fe phase with hyperfine
field about 33\,T and other with a reduced hyperfine field. As our
magnetization measurements on doped samples show presence of a
hard magnetic phase, the phase formed with reduced hyperfine field
may be attributed to it. The values of hyperfine field for such
hard phase are 23\,T for Fe-Al-N and 27\,T for Fe-Zr-N samples.

It may be noted that our XRD, magnetization and CEMS measurements
present a comprehensive information about the structural and
magnetic properties of samples in the as-deposited state and after
annealing at different temperatures. While un-doped sample
undergoes phase formation upon annealing, both local and long
range stability seems to get significantly improved with Al and Zr
doping. However, thermal stability seems to be better with Al than
with Zr. In order to understand the mechanism leading to thermal
stability, we performed self-diffusion measurements of Fe and N in
our samples, which will be discussed in the next section.

\subsection{Self-diffusion measurements}
\label{Sec_Diffusion}

\begin{figure*}\center
\vspace{-1mm}
\includegraphics [width=120mm,height=75mm] {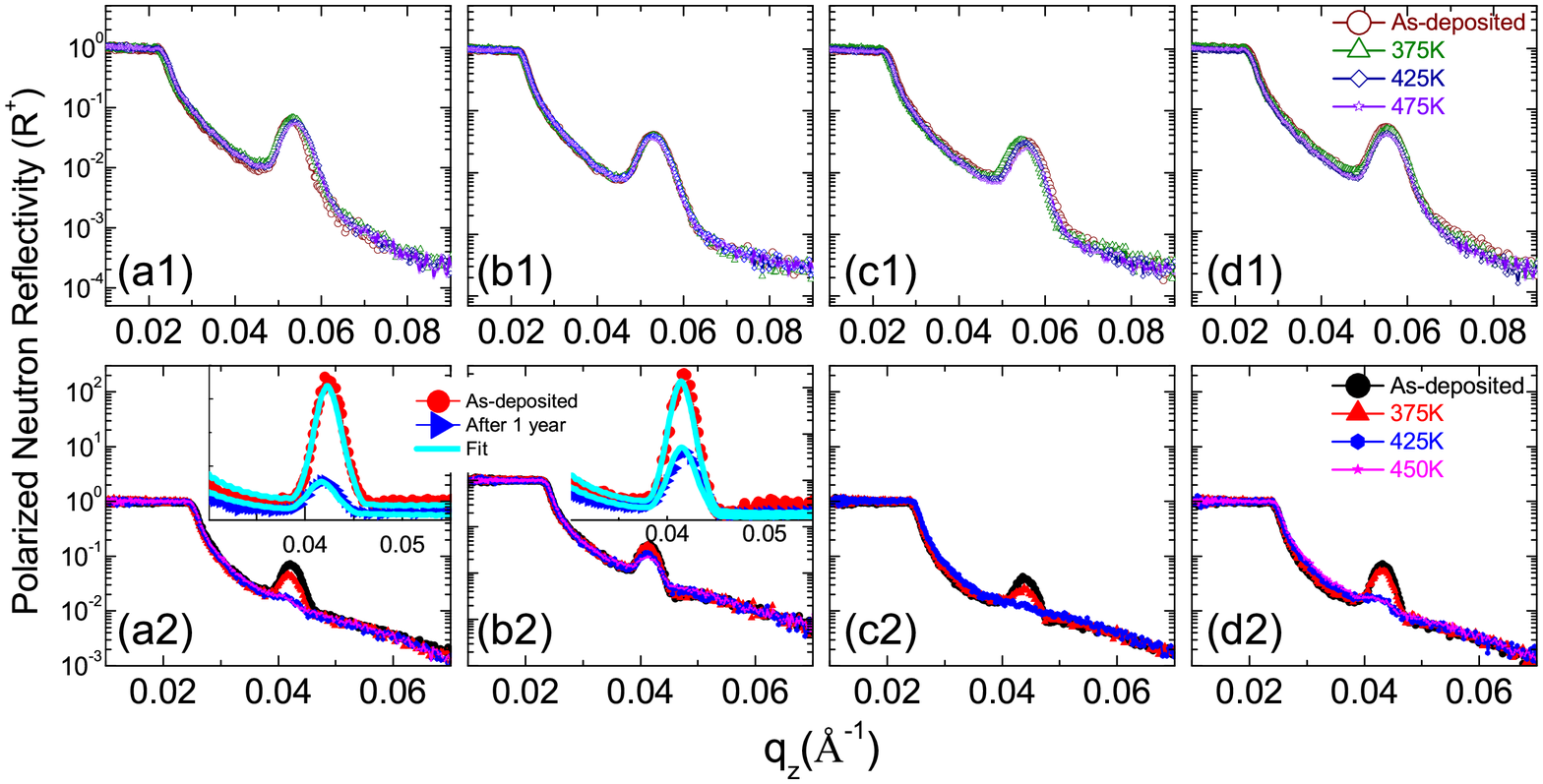}
\vspace {-1mm} \caption{\label{fig:NR}(color online) PNR patterns
of Sub.$\mid$[Fe-X-N$\mid^{57}$Fe-X-N]$_{\times 10}$ for X = 0
(a1), Al (b1), Zr (c1), Ti (d1) and
Sub.$\mid$[Fe-X-N$\mid$Fe-X-$^{15}$N]$_{\times 25}$ for X = 0
(a2), Al (b2), Zr (c2), Ti (d2) in the as-deposited state and
after annealing at various temperatures. Inset of (a2,b2) compares
PNR patterns of Fe-N and Fe-Al-N samples taken just after
deposition and measured after one year.}
\end{figure*}

\begin{figure}\center
\vspace{-1mm}
\includegraphics [width=85mm,height=40mm] {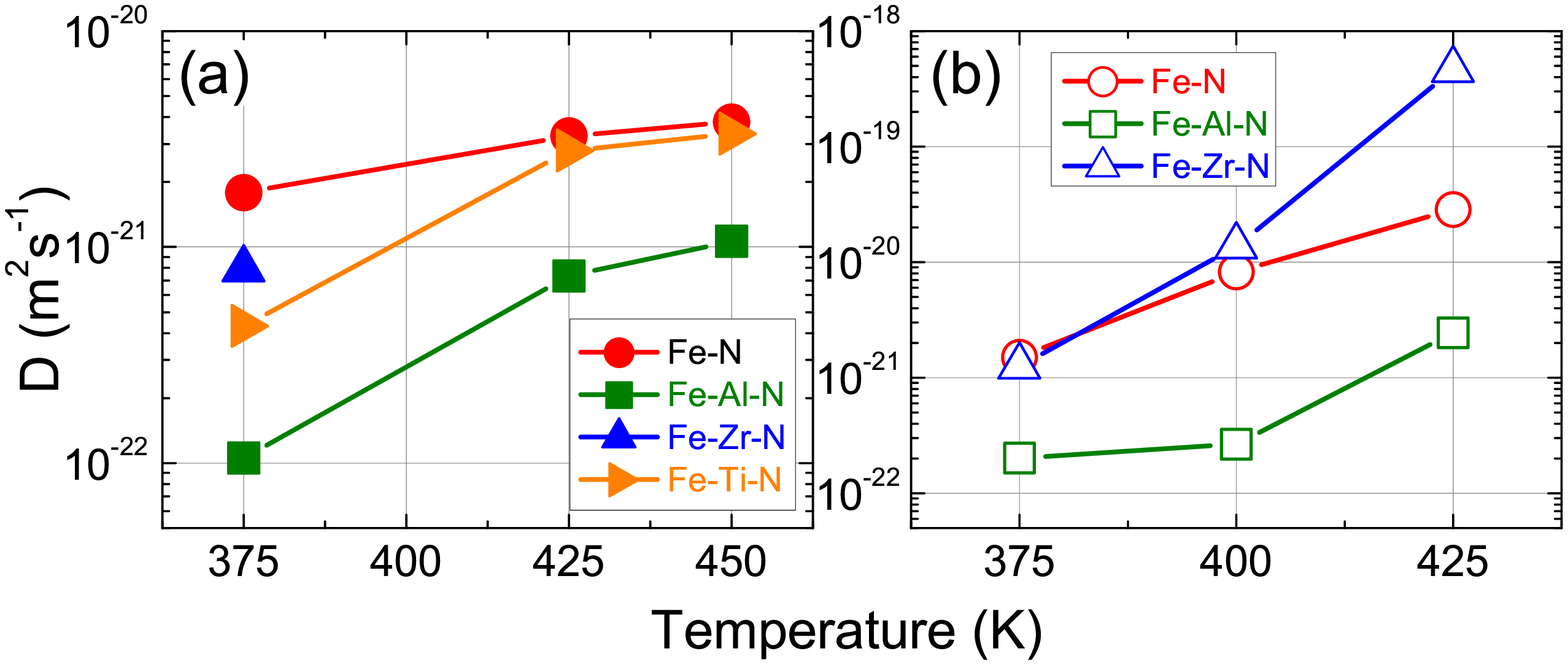}
\vspace {-1mm} \caption{\label{fig:nr_sims_diffusivity}(color
online) Self-diffusivity of nitrogen obtained from  PNR(a) and
SIMS(b) measurements, solid lines are guide to eye. Typical error
bars in the measurements are about the size of scatters.}
\end{figure}

In this section we present self-diffusion measurements performed
using PNR, SIMS and NRR. It is known that these techniques are the
only methods to probe self-diffusion in stable isotopes. While
reflectivity techniques (PNR and NRR) offer an excellent depth
resolution of about
0.1\,nm,~\cite{Spapen:APL80,Greer-JMMM96,PRB:AG:NRR,gupta:JAP2011}
the information obtained from these techniques is `indirect' as
they are based on x-ray/neutron scattering. SIMS on the other hand
provides depth profile of isotopes giving a `direct' information
of diffusivity. Although depth resolution of SIMS is about 5\,nm,
a comparison of reflectivity and SIMS provides reliable
information about self-diffusion. In this work we measured Fe
self-diffusion using PNR and NRR and N self-diffusion using PNR
and SIMS. Complementarities of different techniques was used to
get precise information of Fe and N self-diffusion.

\begin{figure}\center
\vspace{-1mm}
\includegraphics [width=85mm,height=50mm] {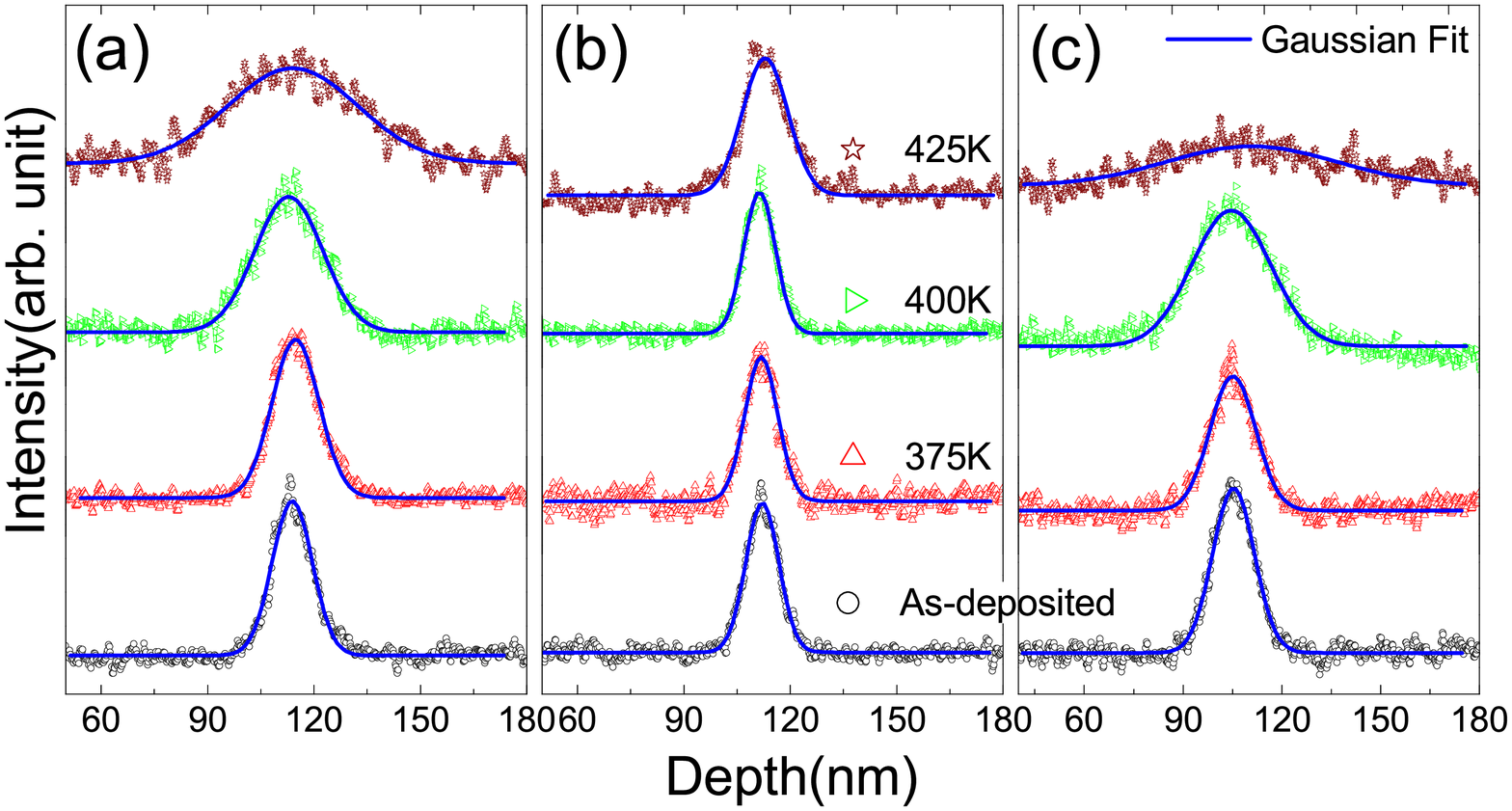}
\vspace {-1mm} \caption{\label{fig:SIMS}(color online) $^{15}$N
SIMS depth profile of
Sub.$\mid$[Fe-X-N(110\,nm)$\mid$Fe-X-$^{15}$N(2\,nm)$\mid$Fe-X-N(110\,nm)]
trilayer samples with X= 0(a), Al(b), and Zr(c) in the
as-deposited state and after annealing at various temperatures.}
\end{figure}

Fig.~\ref{fig:NR} shows PNR patterns of Fe-N(a1,a2),
Fe-Al-N(b1,b2), Fe-Zr-N(c1,c2) and Fe-Ti-N(d1,d2) samples in the
as-deposited state and after annealing at various temperatures. In
fig.(a1-d1) and fig.(a2-d2) a Bragg peak originating due to
scattering length ($b_n$) contrast of $^{57}$Fe/$^{nat}$Fe and
$^{15}$N/$^{nat}$N, respectively. X-ray reflectivity measurements
(not shown) performed on these samples do not show any Bragg peak
due to lack of contrast between isotopes. It confirmed that films
are chemically homogeneous. Since our samples are iron rich and a
large difference between the value of $b_n$ for
$^{nat}$Fe(=9.45\,fm) and $^{57}$Fe(=2.3\,fm) exists, considerably
intense Bragg peak can be seen even with 10 repetition of
bilayers. However, for nitrogen, relatively smaller difference in
$b_n$ for $^{nat}$N(=9.36\,fm) and $^{15}$N(=6.3\,fm) and low
nitrogen concentration ($\sim$11 at.\%), makes it very difficult
to measure nitrogen diffusion. Probably this is the reason that
nitrogen self-diffusion has not yet been reported for magnetic
Fe-X-N thin films. In order to get the appreciable intensity of
Bragg peak, we increased bilayer thickness and the number of
repetitions. From fitting of the PNR data~\cite{SimulReflec}
measured bilayer period for un-doped and Al, Zr or Ti doped
samples are 18.8\,nm, 18.4\,nm, 17.6\,nm, and 17.2\,nm,
respectively. From fig.~\ref{fig:NR}(a2-d2) it can be seen, a
Bragg peak of appreciable intensity due to $^{15}$N/$^{nat}$N
contrast can be clearly observed. Although its intensity is
considerably low (as compared to $^{57}$Fe/$^{nat}$Fe), as
expected.

To study self-diffusion, samples were annealed in a vacuum furnace
at different temperatures for 1\, hour at each temperature. For
$^{57}$Fe/$^{nat}$Fe samples, we find that the intensity of the
peak does not change with annealing (fig.\ref{fig:NR}(a1-d1))
indicating that up to a temperature of 475\,K self-diffusion of
iron is negligible. On the other hand, in $^{15}$N/$^{nat}$N
samples, noticeable nitrogen diffusion can be seen even at a
temperature of 375\,K, and at 425\,K nitrogen gets almost
completely diffused in all but Al doped samples as shown in
fig.~\ref{fig:NR}(a2-d2). Incidentally, we measured un-doped and
Al doped samples kept at room temperature just after deposition
and after about about one year (355 days). Inset of
fig.~\ref{fig:NR} compares PNR patterns of Fe-N(a2) and
Fe-Al-N(b2) samples taken immediately after deposition and after
355 days kept at room temperature. It can be seen that N diffusion
suppresses with Al doping even at room temperature. This is a
clear indication that nitrogen self-diffusion gets remarkably
suppressed with Al doping. In contrast to this, with Zr doping
behavior of nitrogen diffusion is surprisingly unusual, as Zr
doping results in somewhat faster nitrogen diffusion compared to
un-doped Fe-N sample.

From the decay in the intensity of Bragg peak self-diffusivity in
the samples can be calculated using the following
expression:~\cite{Spapen:APL80,Greer-JMMM96,PRB:AG:NRR,gupta:JAP2011}
\begin{equation}\label{NReqn}
\ln\left[\frac{I(t)}{I(0)}\right] = \frac{-8\pi^{2}D t}{d^{2}}
\end{equation}

where $I(0)$ is the intensity of the first order Bragg peak at
time $t = 0$ (before annealing), $d$ is the bilayer thickness and
$D$ is diffusivity. Obtained values of $D$ for Fe-N and Fe-Al-N
samples kept at room temperature for about one year comes out to
be 2$\times 10^{-25} \mathrm{m^2s^{-1}}$ and 1$\times 10^{-25}
\mathrm{m^2s^{-1}}$, respectively. Information about diffusivity
can also be obtained from fitting of PNR data. Following the
fitting procedure as mentioned in section~\ref{Sec_MH}, obtained
values of diffusion length are 3.5\,nm and 2.5\,nm for Fe-N and
Fe-Al-N samples, respectively which are in close agreement with
values obtained using eq.~\ref{NReqn}. It shows that using PNR we
can measure diffusivity precisely down to $1\times
10^{-25}\mathrm{m^2s^{-1}}$. Such small values of diffusivity are
probably the lowest values ever measured.
~\cite{Harald.PRL06,AM:Schmidt:08}

Measured diffusivity at higher annealing temperature gives a
snapshot picture for a fixed annealing time, obtained variation of
$D$ with temperature is shown in
Fig~\ref{fig:nr_sims_diffusivity}(a). As mentioned before, we also
measured nitrogen diffusion using SIMS and for this purpose we
prepared a special trilayer structure
Si(Sub.)$\mid$[Fe-X-$^{nat}$N(110\,nm)$\mid$Fe-X-$^{15}$N(2\,nm)$\mid$Fe-X-$^{nat}$N(110\,nm)],
Such structure is expected to give a peak when looking at $^{15}$N
depth profile. As samples are annealed broadening of this peak
provides information about nitrogen self-diffusion. Applying thin
films solution to Fick's law, the tracer concentration of $^{15}$N
with penetration depth(say $x$) can be expressed
as:~\cite{mehrer2007diffusion}

\begin{equation} \label{SIMSeqn}
c(x,t) = \frac{const.}{2\sqrt{\pi D
t}}exp\left(\frac{-x^{2}}{4Dt}\right)
\end{equation}

Here $t$ is annealing time and $D$ is the diffusion coefficient.
Fig~\ref{fig:SIMS} shows SIMS depth profile of Fe-N, Fe-Al-N and
Fe-Zr-N samples annealed at various temperatures. Here again we
find that as annealing temperature is increased, broadening of
$^{15}$N peak is more for Fe-N and Fe-Zr-N samples as compared to
Fe-Al-N sample. Fitting SIMS profile with a Gaussian function
according to equation~\ref{SIMSeqn}, yields $D$ as:

\begin{equation} \label{SIMS_DR}
D = \frac{\sigma^2_t-\sigma^2_0}{2t}
\end{equation}

Here $\sigma$ is the standard deviation of the Gaussian depth
profile before annealing ($t$=0) and after an annealing time of
$t$. Obtained values of $D$ are plotted in
fig.~\ref{fig:nr_sims_diffusivity}(b). It may be noted that the
depth resolution of SIMS is relatively poor (as compared to
reflectivity). Therefore, absolute values of diffusivity may
differ slightly, still the behavior of nitrogen diffusion is
similar to that obtained with PNR measurements. Since in Zr doped
sample nitrogen diffusion was so fast that it was not possible to
measure it with PNR, in SIMS measurements it can be clearly seen
that with Zr doping nitrogen diffusion become even faster as
compared to the un-doped sample. This is an important result, for
deciding effective dopant in Fe-X-N thin films.

\begin{figure}\center
\vspace{-1mm}
\includegraphics [width=85mm,height=85mm] {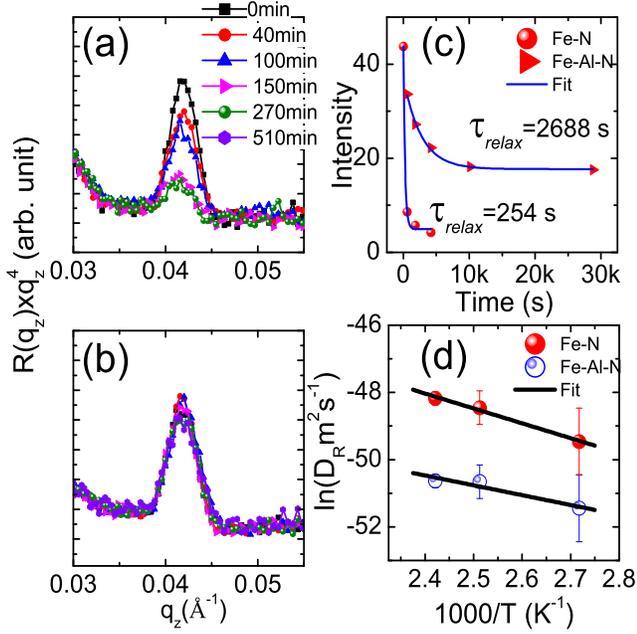}
\vspace {-1mm} \caption{\label{fig:Activation Energy}(color
online) PNR patterns of
Sub.$\mid$[Fe-N(9.4\,nm)$\mid$Fe-$^{15}$N(9.4\,nm)]$_{\times25}$(a)
and
Sub.$\mid$[Fe-Al-N(9.2\,nm)$\mid$Fe-Al-$^{15}$N(9.2\,nm)]$_{\times25}$(b)
samples annealed at 368\,K for different time period. Variation of
Bragg peak intensity with the annealing time for the above samples
annealed at 415\,K(c). Arrhenius behavior of nitrogen diffusion
for the Fe-N and Fe-Al-N samples(d).}
\end{figure}

On the basis of above results, it can be clearly seen that
nitrogen self-diffusion gets reduced only with Al doping. We
therefore also performed PNR measurements with isothermal
annealing experiment in Fe-N and Fe-Al-N samples at 368\,K, 398\,K
and 413\,K for different annealing time. Fig.~\ref{fig:Activation
Energy}(a) and (b) shows PNR patterns of Fe-N and Fe-Al-N samples
annealed at 368\,K for different times. Here intensity was
multiplied by $\mathrm{q_{z}^4}$ to remove decay due to Fresnel
reflectivity. On comparing the reflectivity patterns, it can be
clearly seen that in un-doped sample significant N diffusion
started even at 368\,K which increases further with increasing
annealing time and attains a state of relaxation, whereas in Al
doped sample only marginal diffusion takes place.
Fig.~\ref{fig:Activation Energy}(c) compares variation of
intensity at Bragg peak with time for the Fe-N and Fe-Al-N samples
annealed at 413\,K. It can be seen that intensity decays
exponentially with annealing time. After fitting the data using:
$I(t)=I(0)exp(-t/\tau)$, it was observed that relaxation time
($\tau_{relax}$) for nitrogen diffusion increases by more than an
order of magnitude with Al doping. Such increase in $\tau_{relax}$
provides further insight about involved diffusion mechanism, which
is discussed in section~\ref{DIS}.

It is well known that, in the relaxed state, diffusivities follows
Arrhenius-type behavior given by:~\cite{Faupel_RMP03}
$D_R=D_0\mathrm{exp}(-E/k_BT)$, with $D_R$ is diffusivity in
relaxed state and $D_0$ is the pre-exponential factor, $E$ is
activation energy, $T$ is annealing temperature and $k_B$ is
Boltzman constant. Fig.~\ref{fig:Activation Energy}(d) shows
Arrhenius type behavior of N self-diffusion in Fe-N and Fe-Al-N
thin films. A straight line fit to the data gives value of
$\ln(D_0)$ and $E$. For Fe-N and Fe-Al-N samples values of
($\ln(D_0)$ and $E$) are (-37.39$\pm$1, 0.4\,eV$\pm$0.05) and
(-43.42$\pm$1, 0.25\,eV$\pm$0.02), respectively. It can be seen
that both $\ln(D_0)$ and $E$ decrease with Al doping.

Comparing the value of N diffusivity (e.g. for un-doped sample)
with those reported in literature (for bulk iron nitride), we find
that our values are at least 3-4 orders of magnitude
smaller.~\cite{daSilva:1976} As mentioned already, our samples are
at saturation N concentration, ($\sim$11at.\%), therefore it is
expected that a large fraction of interstitial sites are filled
with N atoms. Therefore, the probability of obtaining neighboring
vacant interstitial sites will get
reduced.~\cite{mehrer2007diffusion}

\begin{figure}\center
\vspace{-1mm}
\includegraphics [width=85mm,height=70mm] {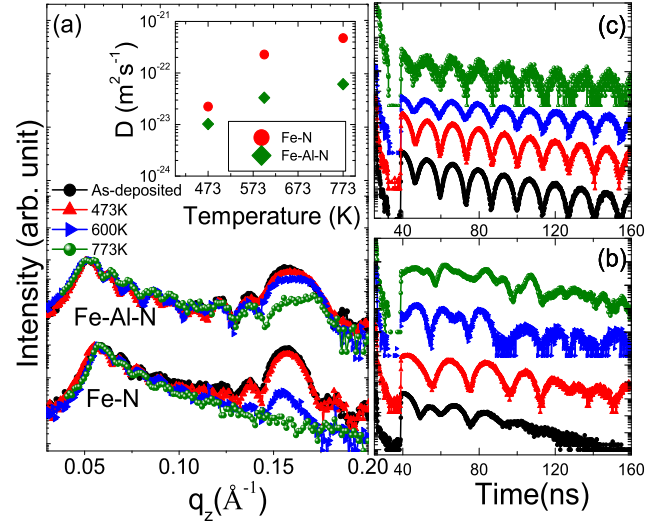}
\vspace {-1mm} \caption{\label{fig:nrr}(color online) NRR patterns
of Sub.$\mid$[Fe-N(2.2\,nm)$\mid^{57}$Fe-N(2.2\,nm)]$_{\times10}$
and
Sub.$\mid$[Fe-Al-N(2\,nm)$\mid^{57}$Fe-Al-$^{15}$N(2\,nm)]$_{\times10}$
samples in the as-deposited state and after annealing at various
temepratures(a), inset of figure shows obtained values of
self-diffusivity. Corresponding NFS patterns of Fe-N (b) and
Fe-Al-N(c) samples.}
\end{figure}

In order to get a complete picture about diffusion in our samples,
it was required that iron self-diffusion should also be measured.
From our neutron reflectivity measurements, we find that there is
no appreciable diffusion up to a temperature of 475\,K. In order
to get a snapshot of diffusion, it was required that bilayer
thickness (13\,nm) should be decreased significantly. We therefore
deposited
substrate$\mid$[$^{nat}$Fe-X-N/$\mid^{57}$Fe-X-N]$_{\times10}$ (X
= 0, Al) samples with bilayer thickness of about 4\,nm under
identical deposition conditions. In such samples Bragg peak in
neutron reflectivity pattern would occur at
$\mathrm{q_z}$=0.16\,\AA$^{-1}$, which is too high to be measured
with neutrons due to limited flux. It is known that for $^{57}$Fe,
nuclear resonance reflectivity (NRR) is a very powerful technique
to get precise information about the
self-diffusion.~\cite{PRB:AG:NRR} In addition when measured in
time domain, nuclear forward scattering (NFS) is Fourier transform
of energy domain reflectivity, that provides direct information
about the local magnetic structure of
Fe.~\cite{Ruffer:PRL:QB:NRR,PRB:NRR:Rohlsberger}

NRR measurements were performed at P01 beamline of PETRA III both
in time integral and time differential modes. For $^{57}$Fe the
lifetime of the excited state of nucleus is about 140\,ns,
therefore 40 bunch mode of PETRA III was used (pulse duration of
about 176\,ns). In time integral mode, both electronic (prompt
with few ns) and nuclear (delayed 40-140\,ns) reflectivities can
be measured simultaneously as they occur in different time windows
and their scattering amplitude is given by : $\mathrm{F=
F_{electronic}+F_{nuclear}}$.~\cite{PRB:NRR:Rohlsberger,PRB:AG:NRR}

NRR measurements in time integral mode were carried out in
$\theta$- 2$\theta$ mode and are shown in fig.~\ref{fig:nrr}(a).
On the other hand, NFS measurements were performed by sitting at
the Bragg peak position and delayed photons were measured in time
differential mode from 40\,ns to 150\,ns (within a bunch). The NFS
spectra show quantum beat pattern spread-over delayed time, which
arises due to interference between various hyperfine field acting
at the resonant nuclei.~\cite{Ruffer:PRL:QB:NRR}

Following a similar approach as mentioned for neutron reflectivity
measurements, Fe self-diffusion was measured by annealing the
samples at different temperatures and performing NRR measurements
subsequently. Fig.~\ref{fig:nrr}(a) shows NRR patterns of Fe-N and
Fe-Al-N samples in the as-deposited state and after annealing
samples at different temperatures. It can be seen that up to a
temperature of 473\,K, Fe diffusion is negligible (also seen from
PNR measurements fig.~\ref{fig:NR}) in both samples. In un-dpoed
sample Fe diffusion is appreciable at 600\,K while at 773\,K it
diffuses completely. Whereas in Al doped sample, only a marginal
diffusion can be seen at 600\,K while at 773\,K Bragg peak can
still be seen. The inset of fig.~\ref{fig:nrr}(a) compares Fe
diffusivity obtained from the decay in intensities of the Bragg
peak using the equation~\ref{NReqn}. As expected, Fe diffusivity
suppress significantly with Al doping. In a recent study Fe
self-diffusion was measured in ion beam sputtered Fe-N samples
prepared using Al and Zr doping and found Fe diffusion suppresses
both with Al and Zr doping.~\cite{JAP:AT}

Corresponding NFS pattern obtained at the Bragg peak position are
given in fig.~\ref{fig:nrr}(b) and (c) for Fe-N and Fe-Al-N
samples, respectively. It can be seen that for Fe-N sample
somewhat complex NFS pattern arises due to interference between
different hyperfine field present in the sample. Whereas in case
of Fe-Al-N sample due to the predominant presence of single
hyperfine field NFS spectra shows uniform quantum beats decaying
with time. Moreover, NFS spectra for Fe-N sample shows variation
with annealing indicating different phase formation, which are
correlated with our XRD and CEMS results discussed in
section~\ref{Sec_XRD_ANN} and~\ref{Sec_CEMS}. In case of Al doped
sample no appreciable change in the NFS pattern can be observed,
it suggests that the local magnetic structure also gets stabilized
with Al doping. These results are consistent with our XRD and CEMS
measurements.

\section{Discussion}\label{DIS}

Combining results discussed above, a picture about the diffusion
mechanism and the influence of dopants on diffusion can be drawn.
We observed that Fe and N self-diffusion takes place in different
temperature regimes. Up to a temperature of about 450\,K, N
diffusion dominates and Fe diffusion is negligible, only above
450\,K considerable Fe diffusion can be observed. Importantly,
both the structure and the magnetic properties remain almost
unchanged even when nitrogen diffuses completely. It indicates
that N diffusion has no significant role in the structural or
magnetic changes in our samples. On the other hand, variation in
the structural and magnetic properties seems to be driven by Fe
self-diffusion which starts above 450\,K. However, dopants have
clear intervention in affecting self-diffusion of both Fe and N.

First, we will discuss the role of dopants on influencing N
self-diffusion. It is known that up to a concentration of
$\sim$11at.\%, N atoms occupies interstitial sites within the Fe
lattice. Numerous studies on self-diffusion behavior of light
elements such as H, C, O, N etc. reveal that they follow
interstitial-type diffusion
mechanism.~\cite{Snoek:1941,PRB:IntDiff:1970,JAP:Zener,JAP:Wert}
In the course of interstitial diffusion, N atoms try to find most
equilibrium interstitial site by crossing a saddle point barrier.
Moreover, interstitial diffusion is strongly affected by pressure
which alters available interstitial volume for diffusing N
atoms.~\cite{Bosman:Pressure:N} In case of un-doped sample, due to
the absence of any impeding force, N diffusion leads to its
redistribution within the lattice. This redistribution favors
nitrogen ordering, as observed from the sharpening in the peak
corresponding to $\alpha^{\prime\prime}$ phase in our XRD results
(fig~\ref{fig:xrd_ann}). As mentioned in ~\ref{Sec_XRD_ANN}, with
Al doping lattice volume decreases by $\sim$0.5\%. This will also
decrease the available interstitial volume for the diffusing N
atoms. Therefore the probability for finding equilibrium
interstitial sites may get reduced, which may result in slower N
diffusion or a larger relaxation time, as observed with Al doping.
On the contrary, lattice volume with Zr doping increases by
$\sim$2\%. In this situation creation of extra equilibrium
interstitial sites may occur, accelerating the diffusion process
of N atoms. Similarly, enhanced diffusion with Ti doping can be
understood.

As pointed out earlier, thermal stability of Fe-X-N thin films are
significantly affected by self-diffusion of Fe. Our results show
that doping of Al has notably reduced the self-diffusion of Fe.
Although, we find that the structure and the magnetic stability
gets improved with both Al as well as Zr doping. It indicates that
the atomic size of dopants does not matter in suppressing Fe
self-diffusion. It seems that the role of dopants on influencing
the self-diffusion of Fe is completely different from N diffusion.
Since $\Delta$H for nitride formation of dopants is low, there is
a large probability that X-N layer may be formed in the grain
boundary region, as observed in some
reports.~\cite{Ding:IEEE:2001,JPCM:97:Wang} This X-N layer may act
as a diffusion barrier for Fe.~\cite{TSF:Nicolet:78} Moreover,
this barrier layer seems to have multiple effects as it not only
suppresses Fe self-diffusion, it also hinders grain growth,
leading to improved soft magnetic properties as observed in our
samples with Al or Zr doping. The improvement in soft-magnetic
properties of Fe-N thin films with dopants was also found in
previous reports.~\cite{RG:JAP12,TSF:Tayal:13,FeTaN:Soft:JAP}
Somewhat superior thermal stability with Al doping can also be
understood from the fact that only with Al doping N diffusion
suppresses.

\section{Conclusion}
\label{Sec_Conclusion}

In the present work we have studied the role of Fe and N
self-diffusion on influencing the structure, magnetic properties
and thermal stability of Fe-X-N thin films. It was observed that
with Al or Zr doping thermal stability gets significantly
improved. Additionally, magnetization measurements revealed that
soft-magnetic properties also gets improved with dopants. To
understand the observed effects, detailed Fe and N self-diffusion
measurements were performed. It was found that dopants have clear
intervention in affecting Fe and N self-diffusion, however, the
mechanism leading to the suppression of Fe and N self-diffusion is
different. In case of N self-diffusion, atomic size of dopants
plays a crucial role. It was observed that N diffusion gets
significantly reduced when the atomic size of dopants is smaller
than that of Fe. A dopant with smaller size lead to compression in
the Fe lattice whereas lattice expansion takes place when a larger
(than Fe) dopant is used. Such lattice distortion caused by
dopants results in alteration of available interstitial volume for
diffusing N atoms. On the other hand Fe self-diffusion gets
suppressed with any dopants, if their  heat of formation is
significantly smaller than that of Fe-N. This happens due to the
formation of a diffusion barrier layer which not only suppresses
self-diffusion of Fe but also hinders the grain growth leading to
improved soft-magnetic properties. In addition it can be concluded
that N diffusion has less significant role (as compared to Fe) in
affecting the thermal instability in Fe-N thin films.

\section*{Acknowledgments}
A part of this work was performed at AMOR, Swiss Spallation
Neutron Source, Paul Scherrer Institute, Villigen, Switzerland and
at P01 beamline, PETRA III, DESY, Germany. We acknowledge
Department of Science and Technology, New Delhi for providing
financial support to carry out PNR and NRR experiments. We
acknowledge V. R. Reddy for CEMS, R. J. Chaudhary for S-VSM and
Layanta Behera for help provided in XRD and SIMS measurements. One
of the authors (AT) thanks to CSIR India for research fellowship.


%

\end{document}